\documentstyle[12pt]{article}

\def\del{\partial}

\def\til{\tilde}
\def\dis{\displaystyle}

\def\a{\rm a}
\def\b{\rm b}

\begin{document}

\vspace*{.7cm}

\begin{center}
{\large\bf A canonical formalism of $f(R)$-type gravity in terms of\\[2mm]
 Lie derivatives}\\[10mm]
\end{center}
\hspace*{15mm}\begin{minipage}{13.5cm}
Y. Ezawa, H. Iwasaki, Y. Ohkuwa$^{\dagger}$, S. Watanabe, N. Yamada and 
T. Yano$^{*}$\\[3mm]
Department of Physics, Ehime University, Matsuyama, 790-8577, Japan\\[1mm]
\hspace*{-1.5mm}$^{\dagger}$Section of Mathematical Science, Department of 
Social Medicine, Faculty of Medicine, University of Miyazaki, Kiyotake, 
Miyazaki, 889-1692, Japan\\[1mm]
\hspace*{-1.5mm}$^{*}$Department of Electrical Engineering, Ehime University, 
Matsuyama, 790-
8577, Japan\\[5mm]
{\small Email :  ezawa@sci.ehime-u.ac.jp, hirofumi@phys.sci.ehime-u.ac.jp,\\
ohkuwa@med.miyazaki-u.ac.jp, shizuka@phys.sci.ehime-u.ac.jp,\\
naohito@phys.sci.ehime-u.ac.jp and yanota@eng.ehime-u.ac.jp}
\\[8mm]
{\bf Abstract}\\[2mm]
 A canonical formalism of $f(R)$-type gravity is proposed, resolving the 
problem in the formalism of Buchbinder and Lyakhovich(BL).
The new coordinates corresponding to the time derivatives of the metric are 
taken to be its Lie derivatives which is the same as in BL.
The momenta canonically conjugate to them and Hamiltonian density are defined 
similarly to the formalism of Ostrogradski. 
It is shown that our method surely resolves the problem of BL.\\


\end{minipage}

\section{Introduction}

Einstein gravity explains the observed universe fairly well.
It has however some theoretical drawbacks, most important of which is the 
problem of initial singulariry\cite{HE,Wald}.
This problem is usually interpreted to imply the limit of applicability of 
Einstein gravity.
There are two possible ways to approach this problem.
One is to modify the theory of gravity in the classical framework.
The most popular modified theory is the higher-curvature gravity(HCG) theories
\cite{Nariai,NT,EM}.
Another is to quantize gravity.
Quantum gravity has a long history but its completion seems to require still 
a long time.

HCG is also required in quantum field theory in curved spacetime\cite{BD}, 
string perturbation theory\cite{Zwie}.
Recently HCG is applied to cosmology to explain, e.g. inflation\cite{Staro} 
or expansion of the present universe\cite{CDTT} which seems to be in the stage 
of an accelerated expansion\cite{Boom,SN,CMB}.
In the former a term proportional to the square of the scalar curvature is 
added to the Lagrangian density of Einstein gravity\cite{Staro,NO} and in 
the latter a term proportional to the inverse of the scalar curvature is 
added\cite{CDTT}.
In the first order Palatini formalism, a wider class of HCG's is also applied 
to cosmology to explain the present expansion of the universe.\cite{1st}\\
Thus the reality of HCG is increasing.

In this work we adress the problem in the canonical formalism of $f(R)$ type 
HCG and propose a consistent formalism.
The curvatures contain the second order time derivatives of the metric which 
in HCG cannot be removed by partial integration as in Einstein gravity.
Thus HCG is a theory with higher order time derivatives(HDT).
Standard procedure for the canonical formalism of HDT has been given by 
Ostrogradski\cite{Ost}.
In the usual method of canonical formalism of gravity in terms of ADM 
variables, this method however is not applicable directly.
The reason is that the curvatures depend on the time derivatives of the lapse 
function and the shift vector so that they obey field equations leading  to 
the breaking of general covariance.

This problem is resolved by the method of Buchbinder and Lyakhovich(BL)\cite
{BL,EKKSY} which generalizes the choice of the new generalized coordinate 
corresponding to the time derivatives of the original generalized coordinates.
Applied to the HCG, the new generalized coordinetes corresponding to the time 
derivatives of the metric was chosen to be the extrinsic curvature.
The time derivatives of the lapse and shift are absorbed in the time 
derivatives of the extrinsic curvature.
However, in this method, a change of the original generalized coordinates 
induces a change of Hamiltonian as will be explained below, which is not 
the case in theories without higer order time derivatives.
For example, in the case of FRW spacetime, the Hamiltonian is different 
whether we use the scale factor or its logarithm as the generalized coordinate.

We propose a canonical formalism which resolves this problem.
We combine the advantageous points of both the method of Ostrogradski and BL.
We choose the extrinsic curvature as the new generalized coordinate as in BL 
and define the momenta canonically conjugate to them similarly to 
Ostrogradski's method.
The extrinsic curvature is the Lie derivative of the three metric which 
reduces to the time derivative in flat space.
The resulting Hamiltonian is shown in fact invariant under the tarnsformations
of the original generalized coordinates which do not change the 3-dimensional 
metric of the hypersurface of constant time.

In section 2, we explain how the Hamiltonian changes under the transformation 
of the original generalized coordinate in the method of BL.
Section 3 is devoted to the presentation of a new canonical formalism for a HCG
of $f(R)$-type.
In section 4 we demonstrate the invariance of the Hamiltonian under the two 
kinds of transformations of the generalized coordinates, the three metric.
Summary and discussion are given in section 5.

\section{A problem in the method of Buchbinder and\\
Lyakhovich}

In this section we show that in the method of BL the Hamiltonian changes under 
the transfornation of the generalized coordinates using a simple model.
Let us consider a system described by a Lagrangian
$$
L=L(q^i,\dot{q}^i,\ddot{q}^i)                                      \eqno(2.1)
$$
in which case the new generalized coordinates are defined as
$$
Q^i\equiv \dot{q}^i.                                               \eqno(2.2)
$$
In the method of BL, the Lagrangian is modified using the Lagrange multiplier 
method so that the definition (2.2) is derived from variational principle.
Denoting the modified Lagrangian as $L^{*}$, it is given as
$$
L^{*}\equiv L+p_{i}(\dot{q}^i-Q^i),\ \ \ L=L(q^i,Q^i,\dot{Q}^i).   \eqno(2.3)
$$
The multipliers $p_{i}$ are the momenta canonically conjugate to $q^i$.
The canonical formalism is obtained by the Legendre transformation starting 
from $L^{*}$.
Momenta $P^i$ canonically conjugate to $Q_{i}$ are given by
$$
P^i\equiv {\del L^{*}\over \del\dot{Q}_{i}}={\del L\over \del\dot{Q}_{i}}.
                                                                   \eqno(2.4)
$$
Then the Hamiltonian $H^{*}$ is given as
$$
H^{*}\equiv p_{i}\dot{q}^i+P_i\dot{Q}^i-L^{*}
=p_{i}Q^i+P_i\dot{Q}^i-L.                                          \eqno(2.5)
$$
Now we make transformations of the generalized coordinates:
$$
\phi^i=f^i(q^j),\ \ \ {\rm or}\ \ \ q^i=g^i(\phi^j).               \eqno(2.6)
$$
Momenta conjugate to $\phi^i$ are denoted as $\pi_{i}$, new generalized 
coordinates as $\Phi_{i}$ and momenta canonically conjugate to them as 
$\Pi_{i}$.
These variables are not uniquely related to old variables.
Since $Q^i$ are related to $\dot{q}^i$ through the variational principle, 
their relations to $\Phi^i$ are not fixed a priori.
The solutions are, however, $Q^i=\dot{q}^i$, it is natural to assume the 
transformation of $Q^i$ to be the same as $\dot{q}^i$.
Differentiation leads to
$$
\dot{q}^i={\del g^i\over \del\phi^j}\dot{\phi}^j\equiv 
{\del g^i\over \del\phi^j}\Phi^j,                                  \eqno(2.7)
$$
so we have the following transformation
$$
Q^i={\del g^i\over \del\phi^j}\Phi^j.                              \eqno(2.8)
$$
Similarly, the momenta $p_{i}$ are Lagrange multipliers, so that their 
transformations are also not fixed a priori.
We fix their transformations by requiring that the modified Lagrangians are 
the same:
$$
L^{*}(q^i,Q^i,\dot{Q}^i)=L^{*}(\phi^i,\Phi^i,\dot{\Phi}^i)
=\bar{L}(\phi^i,\Phi^i,\dot{\Phi}^i)+\pi_{i}(\dot{\phi}^i-\Phi^i). \eqno(2.9)
$$
Then we have
$$
p_{i}(\dot{q}^i-Q^i)=\pi_{i}(\dot{\phi}^i-\Phi^i).                 \eqno(2.10)
$$
Using the first equation of (2.7) and (2.8) in (2.10), we have
$$
\pi_{i}={\del g^j\over \del\phi^i}p_{j}.                           \eqno(2.11)
$$
The Hamiltonian, obtained by the Legendre transformation of the right hand 
side of (2.9), is expressed as
$$
\bar{H}^{*}\equiv \pi_{i}\dot{\phi}^i+\Pi_i\dot{\Phi}^i-\bar{L}^{*}
=\pi_{i}\Phi^i+\Pi_i\dot{\Phi}^i-\bar{L}.                          \eqno(2.12)
$$
Since $L=\bar{L}$, the change of the Hamiltonian is
$$
\Delta H^{*}=\bar{H}^{*}-H^{*}=\Pi_{i}\dot{\Phi}^i-P_{i}\dot{Q}^i. \eqno(2.13)
$$
Now
$$
\Pi_{i}\equiv{\del \bar{L}\over \del \dot{\Phi}^i}
={\del L\over \del \dot{\Phi}^i}=P_{j}{\del g^j\over\del\phi^i}    \eqno(2.14)
$$
and
$$
\dot{\Phi}^i
={\del^2 f^i\over \del q^j\del q^k}\dot{q}^kQ^j
+{\del f^i\over \del q^j}\dot{Q}^j.                                \eqno(2.15)
$$
Using (2.15) and (2.16) in (2.13), we have
$$
\Delta H^{*}={\del^2 f^i\over \del q^k\del q^j}Q^kQ^j\Pi_{i}.      \eqno(2.16)
$$
Thus the Hamiltonian changes under the transformation of the original 
generalized coordinates and reasonable transformation of other canonical 
variables.
If the Hamiltonian represents the energy of the system, which is often the 
case, this is unreasonable physically.
The difference depends on the variables characteristic to higher derivative 
theory.
We note that if the variables transform to make the Hamiltonian invariant, 
the Lagrangian is changed.
In order to resolve this problem, we propose a new canonical formalism which 
can be interpreted to be the generalization of the Ostrogradski's method.

\section{A canonical formalism in terms of Lie derivatives}

The idea is simple.
As noted above, the problem of Ostrogradski's method was resolved by the 
method of BL where the choice of the generalized coordinates corresponding to 
the time derivatives of the original generalized coordinates is extended to 
almost arbitrary functions of these variables.
In the application to gravity, the extrinsic curvature is chosen as the new 
generalized coordinates.
It is the Lie derivatives of the metric which is the generalized coordinates.
Lie derivatives reduce to the time derivatives for flat space.
So this choice can be thought of the simple generalization of the 
Ostrogradski's choice.
Thus we move to the canonical formalism by the Ostrogradski transformation 
contrary to the Legendre transformation with the modified Lagrangian as in 
the method of BL.

Ostrogdadski transformation is carried out as follows.
Let us consider a system described by a Lagrangian containing the $n$-th time 
derivatives of the generalized coordinates $q^i\;(i=1,\ldots,N)$.
$$
L=L(q^i,\dot{q}^i,\ldots,q^{i(n)})                                 \eqno(3.1)
$$
Take the following variation of the action
$$
\delta S\equiv 
\int_{t_{1}+\delta t_{1}}^{t_{2}+\delta t_{2}}L(q^i+\delta q^i,\ldots,
q^{i(n)}+\delta{q}^{i(n)})dt
-\int_{t_{1}}^{t_{2}}L(q^i,\ldots,q^{i(n)})dt                      \eqno(3.2)
$$
where $\delta q^i$ are decomposed as the sum of the variations of the function 
$q^i(t)$, which we denote by $\delta^{*}q^i$, $\delta^{*}q^i\equiv 
(q+\delta q)(t)-q(t)$, and those due to the change of the time , 
$q(t+\delta t)-q(t)=\dot{q}^i\delta t$:
$$
\delta q^i=\delta^{*}q^i+\dot{q}^i\delta t                         \eqno(3.3)
$$
where the second terms are assumed to contribute only near the end points.
Then the above variation is written, retaining only the first order terms in 
small quantities,  as
$$
\delta S
=\int_{t_{1}+\delta t_{1}}^{t_{1}}L(q^1,\ldots, q^{i(n)})dt
+\int_{t_{2}}^{t_{2}+\delta t_{2}}L(q^1,\ldots, q^{i(n)})dt
+\delta^{*}\int_{t_{1}}^{t_{2}}Ldt.                                \eqno(3.4)
$$
Using the approximation
$$
\int_{t_{k}}^{t_{k}+\delta t_{k}}L(q^1,\ldots, q^{i(n)})dt
=L\left(q^1(t_{k}),\ldots, q^{i(n)}(t_{k})\right)\delta t_{k},
\ \ (k=1,2)                                                        \eqno(3.5)
$$
the first two terms in (3.4) become the following
$$
\left[L\delta t\right]_{t_{1}}^{t_{2}}\equiv \delta S_{1}.         \eqno(3.6)
$$
Rearranging the sum in the third term of (3.4), we have
$$
\delta S_{2}\equiv \delta^{*}\int_{t_{1}}^{t_{2}}Ldt
=\left[\delta F\right]_{t_{1}}^{t_{2}}+\delta^{*}S_{2}             \eqno(3.7)
$$
where
$$
\delta F=\sum_{i=1}^N\Biggl[\sum_{s=0}^{n-1}\Biggl\{\sum_{r=s+1}^n
(-1)^{r-s-1}D^{r-s-1}\Bigl({\del L\over \del(D^rq^i)}\Bigr)\Biggr\}
\delta^{*} q^{i(s)}\Biggr]                                         \eqno(3.8)
$$
and
$$
\delta^{*} S_{2}=\sum_{i=1}^N\int_{t_{1}}^{t_{2}}\sum_{s=0}^{n}(-1)^sD^s\Bigl(
{\del L\over \del(D^sq^i)}\Bigr)\delta^{*}q^idt,                   \eqno(3.9)
$$ where $D$ represents the time derivatine, i.e. $\dis D\equiv {d\over dt}$. 
$\delta^{*}S_{2}$ vanishes if we require the variational principle.
The new generalized coordinates are taken as
$$
q_{s}^i\equiv D^sq^i                                               \eqno(3.10)
$$
and the momenta canonically conjugate to these cordinetes are defined by 
the coefficients of the variations of these coordinates in $\delta F$
$$
p_{i}^s\equiv \sum_{r=s+1}^n\left[(-1)^{r-s-1}D^{r-s-1}\Bigl({\del L\over 
\del D^rq^i}\Bigr)\right].                                         \eqno(3.11)
$$
The Hamiltonian is defined as $(-1)\times$ the coefficient of $\delta t$
in $L\delta t+\delta F$ after using (3.3)
$$
H=\sum_{i=1}^N\sum_{s=0}^{n-1}p_{i}^{s}Dq^{i}_{(s)}-L.             \eqno(3.12)
$$
Note that for $s=n-1$, (3.11) has a simple expression
$$
p_{i}^{n-1}={\del L\over \del\dot{q}_{n-1}^i}.                     \eqno(3.13)
$$
Thus the Ostrogradski transformation is a generalization of the Legendre 
transformation.
It is noted that the highest order derivatives need not be the same for all 
$i$.

In this work we investigate the higher-curvature gravity(HCG) of $f(R)$ type 
in which the Lagrangian density is given by a function of the scalar curvature
$$
{\cal L}=\sqrt{-g}f(R)                                             \eqno(3.14)
$$
where $g\equiv \det g_{\mu\nu}$.
The action is given as usual:
$$
S=\int_{t_{1}}^{t_{2}}Ldt,\ \ \ L=\int{\cal L}d^3x                 \eqno(3.15)
$$
We adopt the ADM variables $N({\bf x},t)$, $N^i({\bf x},t)$ and $h_{ij}
({\bf x},t)$ with respect to a hypersurface $t=constant(\Sigma_{t})$ as 
the generalized coordinates. 
In terms of these variables, the scalar curvature is expressed as follows
$$
R
=2N^{-1}h^{ij}\del_{0}K_{ij}-U-2N^{-1}\Delta N
-2N^{-1}(N^k\del_{k}K+2N^{i;j}K_{ij}).                             \eqno(3.16)
$$
Here  $K_{ij}$ is the extrinsic curvature of $\Sigma_{t}$ 
$$
K_{ij}={1\over 2N}\left(\del_{0}h_{ij}-N_{i;j}-N_{j;i}\right),     \eqno(3.17)
$$
and
$$
U\equiv 3K_{ij}K^{ij}-K^2-\til{R},                                 \eqno(3.18)
$$
where $K\equiv h^{ij}K_{ij}$ and $\til{R}$ is the scalar curvature of 
$\Sigma_{t}$.
The determinant $g$ is expressed as $-N^2h$ with 
$h\equiv\det h_{ij}$.

In this case $\delta F$ takes the following form
$$
\begin{array}{ll}
\dis \int d^3x\Biggl[&
\dis {\del {\cal L}\over \del(\del_{0}h_{ij})}\delta^*h_{ij}
-\del_{0}\left({\del {\cal L}\over \del(\del_{0}^2h_{ij})}\right)\delta^*h_{ij}
+{\del {\cal L}\over \del(\del_{0}\del_{k}h_{ij})}\delta^{*}\del_{k} h_{ij}
+{\del {\cal L}\over \del(\del_{0}^2h_{ij})}\delta^{*}\del_{0}h_{ij}
\\[5mm]
&\dis +{\del {\cal L}\over\del(\del_{0}N)}\delta^{*} N
+{\del{\cal L}\over \del(\del_{0}\del_{k}N^i)}\delta^{*} \del_{k}N^i
+{\del L\over\del(\del_{0}N^i)}\delta^{*} N^i\;\Biggr].
\end{array}
$$
From this expression, it appears that $N$ and $N^i$, which represent the choice
 of the coordinate system, have the momenta canonically conjugate to them.
In that case, they are allowed to be only the solutions of dynamical equations 
and the general covariance is broken.
However, the time derivatives, $\del_{0}^2h_{ij}$, $\del_{0}N$, 
$\del_{0}N^i$ and $\del_{0}\del_{k}N^i$ are involved only through 
$\del_{0}K_{ij}$.
It is also noted that time derivatives $\del_{0}h_{ij}$ and 
$\del_{0}\del_{k}h_{ij}$ are involved through $\del_{0}K_{ij}$, $K_{ij}$ and
$\del_{k}K_{ij}$.
Thus $\delta F$ reduces to the integration of
$$
\begin{array}{l}
\dis {\del{\cal L}\over\del(\del_{0}K_{ij})}\delta^{*}K_{ij}
+\dis \Biggl[{\del{\cal L}\over \del(\del_{0}h_{ij})}
-\del_{0}\left({\del{\cal L}\over \del(\del_{0}^2h_{ij})}\right)
-\del_{k}\left({\del{\cal L}\over \del(\del_{0}\del_{k}h_{ij})}\right)\\[7mm]
\hspace*{4cm}\dis -{\del{\cal L}\over \del(\del_{0}K_{kl})}{\del K_{kl}\over 
\del h_{ij}}
+\del_{m}\left({\del{\cal L}\over \del(\del_{0}K_{kl})}{\del K_{kl}\over 
\del(\del_{m}h_{ij})}\right)\Biggr]\delta^{*}h_{ij}
\end{array}                                                        \eqno(3.19)
$$
where we use the relations, e.g.
$$
{\del(\del_{0}K_{kl})\over \del(\del_{0}N)}={\del K_{kl}\over \del N}.
$$
From this expression, if we adopt $K_{ij}$ as the new generalized coordinate, 
instead of $\del_{0}h_{ij}$, time derivatives of $N$ and $N^i$ are absorbed in 
that of $K_{ij}$ and the restriction on $N$ and $N^i$ disappears.
So $K_{ij}$ is taken as the new generalized coordinate as in the method of BL
and is denoted as $Q_{ij}$.
The momenta canonically conjugate to $h_{ij}$ and $Q_{ij}$, $p^{ij}$ and 
$\Pi^{ij}$ respectively, are taken as the coefficients of their variations in 
(3.19), according to Ostrogradski:
$$
p^{ij}={\del{\cal L}\over \del(\del_{0}h_{ij})}
-\del_{0}\left({\del{\cal L}\over \del(\del_{0}^2h_{ij})}\right)
-\del_{k}\left({\del{\cal L}\over \del(\del_{0}\del_{k}h_{ij})}\right)
 -{\del{\cal L}\over \del(\del_{0}K_{kl})}{\del K_{kl}\over 
\del h_{ij}}
+\del_{m}\left({\del{\cal L}\over \del(\del_{0}K_{kl})}{\del K_{kl}\over 
\del(\del_{m}h_{ij})}\right)                                      \eqno(3.20\a)
$$
and
$$ 
\Pi^{ij}={\del{\cal L}\over\del(\del_{0}Q_{ij})}.                 \eqno(3.20\b)
$$
Hamiltonian $H$ is defined as the coefficient of $(-1)\times\delta t$ in 
$L\delta t+\delta F$ after using (3.3).
Then Hamiltonian density is given by
$$
{\cal H}=p^{ij}\del_{0}h_{ij}+\Pi^{ij}\del_{0}Q_{ij}-{\cal L}.     \eqno(3.21)
$$

It is noted that $K_{ij}$ is half the Lie derivative of $h_{ij}$ along 
the normal to the hypersurface $\Sigma_{t}$.
The righthand sides of (3.20) are rewritten explicitely in terms of the Lie 
derivatives as follows
$$
\left\{\begin{array}{l}
p^{ij}=-\sqrt{h}\left[f'(R)Q^{ij}+h^{ij}f''(R){\cal L}_{n}R
+N^{-2}\del_{k}NN^kf'(R)h^{ij}\right]
\\[3mm]
\Pi^{ij}=2\sqrt{h}f'(R)h^{ij}
\end{array}\right.                                                 \eqno(3.22)
$$
where the scalar curvature is expressed as
$$
R=2h^{ij}{\cal L}_{n}Q_{ij}+Q^2-3Q_{ij}Q^{ij}+\til{R}-2\Delta(\ln N).
                                                                   \eqno(3.23)
$$
It is seen from (3.22) that $\Pi^{ij}$ has only the trace part and is written
$$
\Pi^{ij}={1\over d}\Pi h^{ij}\ \ \ {\rm with}\ \ \ 
\Pi=2d\sqrt{h}f'(R)                                                \eqno(3.24)
$$
where $d$ is the dimension of space.
Converting this relation, the scalar curvature is expressed as
$$
R=f'^{-1}(\Pi/2d\sqrt{h})\equiv \psi(\Pi/2d\sqrt{h}).              \eqno(3.25)
$$

\section{Invariance of the Hamiltonian}

In this section, we demonstrate that the Hamiltonian defined in the previous 
section is invariant under the transformation of the generalized coordinates 
induced by (1) the general coordinate transformation on the hypersurface 
$\Sigma_{t}$ and (2) the transformation of the metric $h_{ij}$ to its function 
$G_{ij}(h_{kl})$ such as the transformation of the scale factor $a(t)$ to its 
logarithm $\phi(t)\equiv \ln a(t)$.

\subsection{Invariance under the coordinate transformation on $\Sigma_{t}$}

Let us consider the following general coordinate transformation on $\Sigma_{t}$
$$
\begin{array}{ccl}
x^0&\rightarrow&\bar{x}^0=x^0\\[3mm]
x^i&\rightarrow&\bar{x}^i=f^i(x^1,x^2,x^3).
\end{array}                                                        \eqno(3.26)
$$
Under this transformation, $h_{ij}$ transforms as a second rank tensor, $N$ 
a scalar and $N^i$ a vector.
On the other hand the momenta transform as contravariant tensor densities
$$
\left\{\begin{array}{l}
\bar{p}^{ij}=\dis {\del (x)\over \del (\bar{x})}{\del \bar{x}^i\over \del x^k}
{\del \bar{x}^j\over \del x^l}p^{kl}
\\[5mm]
\bar{\Pi}^{ij}=\dis {\del (x)\over \del (\bar{x})}{\del \bar{x}^i\over\del x^k}
{\del \bar{x}^j\over \del x^l}\Pi^{kl}
\end{array}\right.                                                 \eqno(3.27)
$$
where a overbar represents the transformed quantity and 
$\del (x)/\del (\bar{x})$ is the Jacobian of the transformation.
The transformed Hamiltonian density is defined as in (3.21)
$$
\bar{\cal H}
=\bar{p}^{ij}\del_{0}\bar{h}_{ij}+\bar{\Pi}^{ij}\del_{0}\bar{Q}_{ij}
-\bar{\cal L}.                                                     \eqno(3.28)
$$
The time derivatives and the Lie derivatives do not effect the transformation 
properties so that $Q_{ij}$, $\del_{0}h_{ij}$ and $\del_{0}Q_{ij}$ transform 
as second rank tensors.
Since the Lagrangian density is the scalar density, so is also the Hamiltonian density
$$
\bar{\cal H}={\del (x)\over \del (\bar{x})}{\cal H}.
$$
Therefore we have the desired result
$$
\bar{H}=\int_{t_{1}}^{t_{2}}\bar{\cal H}dtd^3\bar{x}
=\int_{t_{1}}^{t_{2}}{\del (x)\over \del (\bar{x})}{\cal H}dt
{\del (\bar{x})\over \del (x)}d^3x=H.                              \eqno(3.29)
$$
This result is expected since the coordinate system we begin is not 
specified.
Thus what we have shown is essentially the consistency of our method.

\subsection{Invariance under the transformation to the function of the metric}

Next consider the transformation of the generalized coordinates
$$
h_{ij}\rightarrow \phi_{ij}=F_{ij}(h_{kl})\ \ \ {\rm or}\ \ \ 
h_{ij}=G_{ij}(\phi_{kl}).                                          \eqno(3.30)
$$
Under this transformation, three dimensional space is unchanged so that 
the new generalized coordinate $Q_{ij}={\cal L}_{n}h_{ij}/2$ is unchanged
and so is the momentum canonically conjugate to it $\Pi^{ij}
=\del{\cal L}/\del(\del_{0}Q_{ij})$.
These are expressed in terms of the transformed quantities as
$$
\left\{\begin{array}{l}
\dis Q_{ij}={1\over2}{\cal L}_{n}G_{ij}(\phi_{kl})
\\[5mm]
\Pi^{ij}=2\sqrt{h}f'(\psi)G^{ij}(\phi_{kl}).
\end{array}\right.                                                 \eqno(3.31)
$$
Then using (3.22), (3.31) and the relation
$$
\delta h_{ij}={\del G_{ij}\over \del \phi_{kl}}\delta\phi_{kl},    \eqno(3.32)
$$
the first term of $\delta F$, (3.19), takes the following form
$$
-\sqrt{h}\left\{f'(\psi)Q^{ij}
+G^{ij}f''(\psi){\cal L}_{n}\psi
+N^{-2}\del_{k}NN^kf'(R)h^{ij}\right\}{\del G_{ij}\over \del\phi_{kl}}
\delta^{*}\phi_{kl}
+2\sqrt{h}f'(\psi)G^{ij}\delta^{*}Q_{ij}.                          \eqno(3.33)
$$
The momentum $p_{\phi}^{ij}$ canonically conjugate to $\phi_{ij}$ is defined 
as the coefficient of $\delta^{*}\phi_{ij}$:
$$
p_{\phi}^{ij}=-\sqrt{h}\left\{f'(\psi)Q^{kl}
+G^{kl}f''(\psi){\cal L}_{n}\psi
+N^{-2}\del_{k}NN^kf'(R)h^{ij}\right\}{\del G_{kl}\over \del\phi_{ij}}.
                                                                   \eqno(3.34)
$$
From (3.22) and (3.33), we have the following relation
$$
p^{ij}={\del F_{kl}\over \del h_{ij}}p_{\phi}^{kl}.                \eqno(3.35)
$$
Transformed Hamiltonian density ${\cal H}_{\phi}$ is defined as
$$
{\cal H}_{\phi}\equiv 
p_{\phi}^{ij}\dot{\phi}_{ij}+\Pi^{ij}\dot{Q}_{ij}-{\cal L}.        \eqno(3.36)
$$
The change of the Hamiltonian density is given by
$$
\Delta {\cal H}\equiv {\cal H}_{\phi}-{\cal H}
=p_{\phi}^{ij}\dot{\phi}_{ij}-p^{ij}\dot{h}_{ij}.                  \eqno(3.37)
$$
Using (3.32) and (3.34), it is shown that $\Delta {\cal H}$ vanishes.
This result should be necessary if the Hamiltonian has something to do with 
the energy.

Application to the FRW spacetime, in which case the transformation is from 
the scale factor $a(t)$ to its function $G(a)$, e.g., $\ln a(t)$, is 
straightforward.

\section{Summary and discussions}

We proposed a canonical formalism of HCG by combining those of Ostrodradski
and BL using the Lie derivatives instead of the time derivatives.
Invariance of the Hamiltonian is shown under the transformation of generelized 
coordinates preserving the hypersurface $\Sigma_{t}$ which is lacking in BL.
In fact, the transformation properties in HCG have not been addressed.

The result is important in HCG where the conformal transformation to Einstein 
frame is an often used technique\cite{equiv}.
The problem with this transformation is which of the metric is physical, i.e.
the observed one.
Many criteria have been proposed to select the physical one, however 
the problem has not been settled\cite{MS}.
However the conformal transformation depends on the curvature, so does on 
the momenta from the viewpoint of canonical formalism.
It is not obvious whether the canonical equations of motion in both frame is 
equivalent.
Also it would be necessary to show that the Poisson brackets defined in each 
frame are consistent.
If the answers to these problems are negative, equivalence of both frames is 
broken and the canonical quantization leads to different quautum theories.
The results of these problems will be reported separately.

Finally we comment on quantum gravity.
It is well known that there are two versions.
One is the quantum theory of gravitons based on the duality of gravitational 
wave and graviton and has the same footing as the ordinary quantum theory of 
matter and radiation.
This version, however, requires a background spacetime.
If the quantum theory is the fundamental theory in physics, the background 
spacetime should also be determined quantum theoretically.
It is the subject of quantum cosmology which is the second version of quantum 
gravity.
In this version no guiding principle such as duality of particles and waves 
has been known.
The usually adopted  procedure is the canonical quantization.
For this, the classical theory is a prerequisite.
Such a theory may not be the Einstein gravity but a kind of HCG.
If this were the case, investigations of the early stage of the univerese, or 
string theory might give some clues.
The strings feel tidal force, which necessarily leads to HCG.
Observational informations are generally more important.
These informations may be brought about from future cosmological observations.

\end{document}